\documentclass{article}
\usepackage[T1]{fontenc}
\usepackage[utf8]{inputenc}
\usepackage[margin=1.2in]{geometry}
\usepackage[dvipsnames]{xcolor} 
\usepackage{amsmath,amssymb,amsthm}
\usepackage{microtype}
\usepackage[numbers,sort&compress]{natbib}
\usepackage[hidelinks,bookmarksnumbered]{hyperref}
\hypersetup{pdftitle={Exact Agreement Expansion up to the Top via Non-polarity},pdfauthor={Tali Kaufman and Huy Tuan Pham}}

\theoremstyle{plain}
\newtheorem{thm}{Theorem}
\newtheorem{lem}[thm]{Lemma}
\newtheorem{cor}[thm]{Corollary}
\newtheorem{claim}[thm]{Claim}
\theoremstyle{definition}
\newtheorem{defn}[thm]{Definition}
\theoremstyle{remark}

\newcommand{\Ex}{\mathbb E}
\newcommand{\Var}{\operatorname{Var}}
\newcommand{\Prob}{\mathbb P}

\title{Exact Agreement Expansion up to the Top via Non-polarity}
\author{Tali Kaufman\thanks{Supported by ISF and the Israeli interdisciplinary center for quantum information.}\\
  {\small Department of Computer Science}\\
  {\small and Artificial Intelligence}\\
  {\small Bar-Ilan University}
  \and
  Huy Tuan Pham\thanks{Supported by a Clay Research Fellowship, Sloan Research Fellowship, NSF CAREER award DMS-2543870, and the Bell System Fellowship at the Institute for Advanced Study.}\\
  {\small Department of Mathematics}\\
  {\small University of Chicago}}
\date{}

\begin{document}
\maketitle

\begin{abstract}
Agreement expansion asks whether local assignments to faces of a poset that usually agree on a sampled common face must agree with a single global function on most faces. Agreement expansion is {\em exact} ({\em approximate}) if local assignments need to exactly (approximately) agree with the global function. We suggest a new proof technique for agreement expansion that enables agreement expansion in settings which were previously beyond reach, such as:  exact agreement expansion for Grassmannian posets (before only simplicial complexes support exact agreement expansion); agreement expansion based on assignments to top faces of the a complex (before agreement expansion is generally proven only for assignments to low dimensional faces);  agreement expansion from one-sided spectral expansion (before, two sided spectral assumption was needed in general settings).

Specifically, for down-closed simplicial and Grassmannian complexes, we prove an exact agreement theorem between any two layers assuming one-sided spectral expansion in links of singletons. As an application, we show that full bases of every rank-$k$ matroid admit exact agreement with soundness $O(k^2/[h(k-h)])$ when tested through $h$-element independent sets. We also obtain exact agreement on Grassmannian posets under one-sided spectral expansion, with stronger soundness under a local distance guarantee. Our key proof ideas center around non-polarity of the distributions of local assignments, probed via a random exposure of elements.
\end{abstract}

\section{Introduction}

Agreement expansion is a local-to-global property of a family of sets. Given an assignment to each set, an agreement test samples two sets containing a common smaller set and checks that their assignments agree there. The family is an \emph{exact agreement expander} if a small rejection probability implies that most local assignments are restrictions of one global assignment. In approximate agreement, errors on a small fraction of the elements are also allowed within each local assignment.

Agreement tests first arose in low degree testing~\cite{AS97,RS96,RS97}, and have important applications to PCPs and locally testable codes. Dinur and Steurer~\cite{DS14} established exact agreement in the high-acceptance regime for direct product tests. The connection with high dimensional expansion established by Dinur and Kaufman~\cite{DK17} gave agreement tests on sparse families of sets. Subsequent works have extended the sufficient conditions for agreement testing~\cite{DD19,KM20}. 

In this work we develop a general framework for exact agreement from spectral gaps of random walks on the complex. It applies to several settings, including sparse simplicial and Grassmannian complexes with one-sided spectral expansion. Our framework supports assignments to top dimensional faces of the poset, where previous work generally assumes assignments to faces of lower dimensions. Moreover, we prove stronger bounds for agreement expansion if the assignments come from a small code with large distance. Below we give an overview of our results before going to further formal details in Section \ref{subsec:intro-statements}. 

\paragraph{Agreement tests and soundness.}
In the general setup, let $V$ be a finite ground set and let $\mathcal K$ be a family of $N$-element subsets, equipped with a distribution $\mu$. Each $K\in\mathcal K$ has a local assignment $F_K:K\to\Sigma$, where $\Sigma$ is a finite alphabet. We consider a downward sampling rule which chooses an $n$-element set $H\subset K$. The agreement test samples $H$ from this marginal and then two independent parents $K,K'$ conditioned on $H$. Define
\[
 \tau=\Pr[F_K|_H\ne F_{K'}|_H]
 \]
 as the rejection probability of the test. We then let
 \[
 \Delta=\min_{g:V\to\Sigma}\Pr_K[F_K\ne g|_K]
\]
denote the distance of the local assignments to the (optimal) global assignment. 
The soundness constant of the agreement test is the least $C$ such that $\Delta\le C\tau$ for every ensemble of assignments. Its agreement expansion constant is the inverse of the soundness.

\paragraph{Agreement expansion up to the top.}
A central question is whether agreement can be tested on the top faces of the poset. Results that require additional layers above the assignment layer do not directly address this question. Exact agreement on top faces is known for two-sided local spectral expanders and for one-sided partite complexes~\cite{DD19}. Kaufman and Mass~\cite{KM20} give a local-to-global theorem under one-sided expansion and an agreement hypothesis on vertex links.

Our result gives agreement expansion up to the top for general one-sided local spectral expanders, without these additional structural assumptions, which permits new applications not covered by earlier works. %

\paragraph{Exact agreement from one-sided spectral expansion.}
For each element $x$, consider the down-up walk on the sets $K\in\mathcal K$ containing $x$: sample $H\subseteq K$ from the prescribed distribution conditioned on $x\in H$, and then sample $K'$ conditional on $H$. Suppose each of these walks has spectral gap at least $\gamma>0$.

Assume that, for $\rho=n/N\in(0,1)$ and every top face $K$,
\[
 \mathbb{P}(x\in H\mid K)=\rho \,\, \forall x\in K, \qquad \operatorname{Var}(|E\cap H|\mid K)
 \le\rho(1-\rho)|E| \,\, \forall E\subseteq K.
\]
These assumptions hold for uniform subface sampling in every down-closed simplicial complex, and in every down-closed Grassmannian complex with projective or affine points as coordinates. For instance, the assumptions hold in a simplicial complex with $|K|=N$ and $|H|=n$.

Under these assumptions, our main result, Theorem~\ref{thm:regularized-exact}, proves
\[
 \Delta\le
 \left(\frac8\gamma+\frac{64(1-\rho)}{\rho\gamma^2}\right)\tau.
\]
Thus, for down-closed simplicial and Grassmannian complexes, spectral gaps of random walks in links of singletons give exact agreement without requiring any layers above the assignment layer.

\paragraph{Full bases of arbitrary matroids.}
As an immediate corollary, our main result implies exact agreement expansion for top-dimension faces under one-sided local spectral expansion, via known local-to-global bounds for high order walks~\cite{KO18,AL20}. In particular, let $\alpha_i$ bound the second eigenvalue of the non-lazy walk in every link of an $i$-element face in a simplicial complex. If $0\le\alpha_i\le a<1$ and $\sum_i\alpha_i\le A$, for fixed $a,A$, Corollary~\ref{cor:one-sided-agreement-expansion} gives soundness
\[
 O_{a,A}\left(\frac{k^2}{h(k-h)}\right),
\]
for the agreement test between faces of size $k$ and $h$. In particular, the soundness is $O_{a,A,\varepsilon}(1)$ for $\varepsilon k\le h\le(1-\varepsilon)k$, and $O_{a,A}(k)$ for consecutive layers $h=k-1$.

The independence complex of a matroid has one-sided local spectral bounds $\alpha_i=0$~\cite{ALOV19}. Applying the theorem proves exact agreement for the uniform distribution on full bases of every rank-$k$ matroid. When tested through induced $h$-element independent sets, the soundness is
\[
 O\left(\frac{k^2}{h(k-h)}\right),\qquad 1\le h<k.
\]
This gives $O_\varepsilon(1)$ soundness when $\varepsilon k\le h\le(1-\varepsilon)k$, and $O(k)$ when $h=k-1$. Earlier results treat truncations of arbitrary matroids and full bases of certain linear matroids~\cite{DD19}. The conclusion here applies to full bases of every matroid.

\paragraph{Exact agreement on Grassmannian posets.}
Our work also gives {\em exact} agreement expansion between any two layers of {\em sparse} Grassmannian posets under one-sided spectral expansion. For assignments to $k$-dimensional spaces tested through $h$-dimensional spaces, the soundness is $O(q^{k-h})$ when the required spectral gaps are bounded below, where $q$ is the field size. In particular, we obtain exact agreement on two consecutive layers with soundness $O(q)$.

Earlier work of Dikstein and Dinur~\cite{DD19} gives approximate agreement over complete Grassmannians under a local distance guarantee and a separation between the two layers. Exact agreement is also known for assignments from suitable lifted affine-invariant codes~\cite{DHKR22}. Our theorem gives exact agreement for arbitrary local assignments and applies to sparse families satisfying the spectral assumptions. We further obtain constant soundness for approximate agreement without a local distance guarantee.

\paragraph{Stronger agreement under a local distance guarantee.}
If we consider posets where the assignments come from local codes whose distinct restrictions have large distance on every tested faces then we get stronger soundness of agreement expansion. Specifically, when we consider assignments to subspaces of the Grassmannian complex that come from a code with a large distance, the soundness of exact agreement can be improved from $O(q^{k-h})$ to $O(q)$. Thus the soundness no longer grows with the separation between the two layers.
For the general improved agreement expansion when assignments come from a code see Theorem   \ref{thm:agreement-code}.

\paragraph{Applications to locally testable codes.}
Our exact agreement theorem is a key ingredient in the companion work of Liu and Pham~\cite{compa}, which constructs locally testable codes from sparse Grassmannian posets. The construction starts with a family of local codes, one on each $k$-dimensional space in the poset. The lifted code consists of global assignments whose restriction to every such space belongs to the corresponding local code. Suppose that each local code is locally testable by sampling an $h$-dimensional subspace. Under the spectral expansion and local-code assumptions of~\cite{compa}, the lifted code is then locally testable by these same local tests. Thus local testability on the $k$-dimensional spaces gives global local testability using only the subspaces in the sparse poset.

Exact agreement is crucial in this argument: it gives a global assignment whose restrictions coincide with the local codewords on most $k$-dimensional spaces, and hence satisfy the corresponding code constraints exactly. This is the consistency step in the companion paper's local-to-global argument. Our theorem applies between any two layers, including consecutive layers, so this step does not require a large separation between $h$ and $k$. Moreover, under the local distance guarantee, the agreement soundness is $O(q)$, independently of $k-h$. Both the freedom to use nearby layers and this quantitative bound are important for passing from the local tests to a test of the lifted code.

\subsection{Setup and statements of results}
\label{subsec:intro-statements}

We now define the setup and state our main results.

\paragraph{The agreement test and its associated walks.}
Let $\mathcal K$ be a family of $N$-element subsets of a finite set $V$, with distribution $\mu$, and let $D(H\mid K)$ be a rule for sampling an $n$-element subset $H$ of $K$, where $0<n<N$. Let $\mu_H$ denote the induced distribution on the $n$-element subsets $H$ obtained by sampling $K\sim\mu$ and then $H$ according to $D(H\mid K)$. Let $U(K\mid H)$ denote the conditional distribution of $K$ given $H$, so that
\[
 U(K\mid H)=\frac{\mu(K)D(H\mid K)}{\mu_H(H)}.
\]
The agreement test first samples $H\sim\mu_H$, then samples $K,K'$ independently from $U(\cdot\mid H)$, and compares $F_K|_H$ and $F_{K'}|_H$. Its rejection probability $\tau$ and the distance $\Delta$ from a global assignment are as defined above.

Let $\rho=n/N$. We assume that each element of a top face is sampled with the same probability, and that the following variance bound holds: for every $K$ of positive mass,
\[
 \begin{aligned}
 \Pr[x\in H\mid K]&=\rho &&\text{for every }x\in K,\\
 \operatorname{Var}(|E\cap H|\mid K)&\le\rho(1-\rho)|E|
       &&\text{for every }E\subseteq K.
 \end{aligned}
\]
These assumptions hold for uniform subface sampling in every downward-closed simplicial or Grassmannian poset. Indeed, symmetry within $K$ gives equal inclusion probabilities for all elements and $\Pr[x,y\in H\mid K]=n(n-1)/(N(N-1))$ for distinct elements. Hence, writing $m=|E|$,
\[
 \operatorname{Var}(|E\cap H|\mid K)
 =\rho(1-\rho)m\frac{N-m}{N-1}
 \le\rho(1-\rho)m.
\]

For an element $x$, the {\em down-up walk in the link of $x$} starts at a top face $K$ containing $x$, samples $H\subset K$ according to $D(\cdot\mid K)$ conditioned on $x\in H$, and then samples $K'$ according to $U(\cdot\mid H)$. Denote its transition operator by $M_x$. It is reversible with stationary distribution
\[
 \mu_x(K)=\mu(K\mid x\in K).
\]

\paragraph{Spectral gap.}
Let $\gamma_x = 1 - \lambda_2(M_x)$ be the spectral gap of this random walk, so that for every real-valued function $f$ on the top faces containing $x$,
\[
 \frac12\mathbb E_{K\sim\mu_x,\ K'\sim M_x(K,\cdot)}
       (f(K)-f(K'))^2
 \ge\gamma_x\operatorname{Var}_{\mu_x}(f).
\]
Let $\gamma>0$ be a common lower bound on $\gamma_x$ over elements of positive marginal mass. In a simplicial complex, this is the $(h-1,k-1)$ down-up walk in the link of $x$, when top faces have $k$ elements and tested faces have $h$ elements.

We are now ready to state our main result.
\begin{thm}[Main theorem; see Theorem~\ref{thm:regularized-exact}]
\label{thm:intro-regularized}
In the setup above, suppose that all the walks $M_x$ have spectral gap at least $\gamma>0$. Then, for every finite alphabet and every collection of local assignments, the exact agreement test has soundness at most
\[
 C(\rho,\gamma):=\frac8\gamma+\frac{64(1-\rho)}{\rho\gamma^2}.
\]
\end{thm}
The theorem applies directly to the top layer. Applying known bounds on the walks in singleton links gives the consequences for one-sided simplicial expanders and full matroid bases discussed above.

In particular, in a one-sided local spectral expander, i.e. a simplicial complex where the second eigenvalues of the non-lazy walks in links of $i$-element faces are bounded by $\alpha_i$, with $0\le\alpha_i\le a<1$ and $\sum_i\alpha_i\le A$, the agreement test between the $k$ and $h$-element faces has soundness
\[
 O_{a,A}\left(\frac{k^2}{h(k-h)}\right).
\]

\paragraph{Grassmannian posets.}
Let $P$ be a down-closed family of linear or affine subspaces of a fixed ambient space over $\mathbb F_q$, with all maximal spaces of dimension $k$. A distribution on these top spaces induces the lower distributions by choosing a uniform subspace of the desired dimension. We test assignments on the $k$-spaces through $h$-spaces, where $1\le h<k$. Here $k,h$ are dimensions of the relevant subspaces. 

In the affine case, assignments are functions on the points of each affine space. In the linear case, we consider assignments on all vectors, as well as assignments on projective points, which are one-dimensional subspaces. Write
\[
 n_j=\frac{q^j-1}{q-1}.
\]
A linear $j$-space has $n_j$ projective points, while an affine $j$-space has $q^j$ points. In either point model, the uniform subspace sample satisfies the variance assumptions of the main theorem. %

For assignments on all vectors of a linear space, we fix an ordering of the $q-1$ nonzero vectors on each projective line, and group their labels into one element of $\Sigma^{q-1}$. Exact agreement on these coordinates is precisely agreement on all nonzero vectors, and the corresponding walks are the projective-point-conditioned walks. The origin belongs to every tested subspace. We therefore also use the unconditioned down-up walk, which samples $H\subset K$ and then $K'$ above $H$ without fixing an element. Denote its spectral gap by $\gamma_0$.

\begin{cor}[Exact agreement on Grassmannians; see Corollary~\ref{cor:grassmannian-exact}]
\label{cor:intro-grassmannian}
Suppose the walks conditioned on an element have spectral gaps at least $\gamma>0$. The exact $(k,h)$-agreement test has soundness at most $C(\rho,\gamma)$, where
\[
 \rho=\frac{q^h-1}{q^k-1}\quad\text{for projective points},
 \qquad
 \rho=q^{h-k}\quad\text{for affine points}.
\]
For assignments on all vectors of linear spaces, the soundness is at most $C(\rho,\gamma)+1/\gamma_0$, with the projective value of $\rho$, provided $\gamma_0>0$.
\end{cor}
In particular, when the required spectral gaps are bounded below by positive constants, the soundness is $O(q^{k-h})$. This gives $O(q)$ for consecutive layers and $O(q^{k/2})$ for $h=k/2$.

One-sided local spectral expansion provides a sufficient condition for the required spectral gaps in a linear Grassmannian. For each subspace $S$ of dimension at most $k-2$, consider the walk on one-dimensional spaces in the quotient link above $S$, where the edge between distinct $L,L'$ has weight proportional to the induced measure of their span $L+L'$. Let $\alpha$ bound the second largest eigenvalue of all these walks. Theorem~\ref{thm:grassmannian-walk}, from the companion paper~\cite{compa}, shows that $\alpha<c/k$ for a sufficiently small absolute constant $c$ gives uniformly positive spectral gaps for the down-up walks used here. %

We also show that we can obtain substantially better bounds if exact agreement is weakened to approximate agreement.
\begin{cor}[Approximate agreement; see Section~\ref{sec:approximate-agreement}]
\label{cor:intro-approximate}
Suppose the walks conditioned on projective or affine points in Corollary~\ref{cor:intro-grassmannian} have spectral gaps bounded below by a positive constant. There is a global function $g$ such that
\[
 \Pr_K\bigl[|\{x\in K:F_K(x)\ne g(x)\}|\ge\delta|K|\bigr]
 \le O(\tau)
\]
whenever
\[
 \delta\ge\frac{c}{n_h}\quad\text{in the projective or all-vector linear model},
 \qquad
 \delta\ge cq^{-h}\quad\text{in the affine model},
\]
for any fixed $c\ge1$. Here $|K|$ denotes the number of elements on which $F_K$ is defined. The implicit constant depends only on $c$ and the spectral gap bound. %
\end{cor}

\paragraph{Assignments from local codes.}
Suppose each $F_K$ belongs to a local code $\mathcal C_K\subseteq\Sigma^K$. The {\em local distance guarantee at rank $h$} means that, for every tested face $H$, any two distinct assignments in
\[
 \{f|_H:K\supseteq H,\ f\in\mathcal C_K\}
\]
differ on at least a $\theta$ fraction of $H$.

\begin{cor}[Improved exact agreement; see Corollary~\ref{cor-agreement-code}]
\label{cor:intro-local-distance}
Let $P$ be a down-closed linear Grassmannian, with assignments on all vectors and local distance at least $\theta$ on the tested $h$-faces, where $0<\theta\le1$. Suppose
\[
 q^{h-1}\ge1+\frac{8(q-1)}\theta.
\]
For each $s\in\{h,k-1\}$, assume that the unconditioned down-up walk through $s$-faces and all such walks conditioned on a projective point have spectral gaps at least $\gamma_*>0$. Then the exact $(k,h)$-agreement test has soundness $O_{\gamma_*}(q)$.
\end{cor}

\subsection{On our non-polarity based proof technique}
\label{subsec:intro-proof}

Our proof centers around two ideas: {\em non-polarity} of the induced distributions on tested faces, and a random exposure argument to detect deviations from the global majority function. At each element, this function takes a most frequent label, with ties broken arbitrarily. Let $\pi_H$ be the distribution of $F_K|_H$ for a random parent $K$ of $H$. Its non-polarity is
\[
 \Psi(H)=\Pr_{f,f'\sim\pi_H}[f\ne f'],
\]
where $f,f'$ are independent. Thus $\tau=\mathbb E_H\Psi(H)$. We call a distribution polar when most of its mass lies on a single assignment. For technical reasons, we will be working with a regularized version of non-polarity. We then lower bound the non-polarity measure by the distance from the global majority, while upper bound the non-polarity by the rejection probability of the agreement test. 

To obtain these relations, we use a random exposure argument. We give each element an independent exponential clock of rate one and add an independent stopping clock of rate $a$. Expose the elements in clock order until the first disagreement with the majority function or the stop. If an assignment has $m$ errors, each error is recorded with probability $1/(m+a)$. Our regularized non-polarity measure is constructed from the distribution of the first disagreement obtained in this exposure process. By using the one-sided spectral expansion, we can lower bound the non-polarity by the distance from the global majority. 

We then compare this exposure with one performed inside the tested face, with stopping rate $\rho a$. Two assignments that agree on the tested face give identical local witness distributions, so rejection probability of the test bounds from above a local version of non-polarity. We then show that the local exposures inside tested faces give a good approximation to the global exposures, thus giving an upper bound on the non-polarity via the rejection probability of the agreement test. By choosing the stopping rate appropriately, we then obtain the agreement bound. The full argument is given in Section~\ref{sec:agreement}. %

Previous methods to derive agreement testing typically use the small rejection probability of the agreement test to define various local global functions that mostly agree with the assignments in their vicinity. Then they use sampling properties to {\em glue} the local global functions into a single global function that mostly agree with local all views.  Thus, previous methods rely on few disagreements to find a global function that is mostly agreeing with the various local views. To run this approach, one usually needs to require further assumptions (e.g. additional layers above in the complex). %

\paragraph{Organization.} 
After introducing some preliminaries in Section \ref{sec:preliminaries}, we prove the main theorem, giving the soundness of the agreement test under one-sided spectral expansion, in Section~\ref{sec:agreement}. In Section~\ref{sec:applications}, we utilize local-to-global results to obtain agreement testing corollaries in local spectral expanders, yielding applications on bases of matroids and Grassmannian posets. Finally, in Section~\ref{sec:local-distance}, we show a general result yielding improved agreement expansion under local distance guarantee, with application to the Grassmannian setting. 

\paragraph{AI Statement.}
ChatGPT was used for proofreading, minor edits and improvement in exposition. 

\section{Preliminaries}\label{sec:preliminaries}

In this section, we introduce the formal setup and notations for use in the rest of the paper.

Throughout, $X$ is a finite ground set and $\Sigma$ is a finite alphabet. For functions $f,g$ on the same domain, write
\[
 d(f,g)=|\{x:f(x)\ne g(x)\}|.
\]
We write $f\ne_\delta g$ when $d(f,g)\ge\delta|\operatorname{dom}(f)|$. All probabilities and conditional expectations are taken with respect to the measures specified below; faces of zero mass are omitted.

\subsection{Agreement and local distance}

An assignment instance is a collection $\mathcal F=(F_K)_{K\in P_k}$, where $F_K:K\to\Sigma$. The $(k,h)$-agreement test samples $H\sim\mu_h$, then independent parents $K,K'$ from $\mu_k$ conditional on containing $H$, and checks equality on $H$. Its rejection probability is
\[
 \tau_h=\Prob[F_K|_H\ne F_{K'}|_H].
\]
We omit the subscript when the tested layer is fixed. The sampled face $H$ may be a proper subset of $K\cap K'$.

\begin{defn}[Agreement expansion]
The exact $(k,h)$-agreement test has soundness $C$ if every instance admits a global function $g:X\to\Sigma$ such that
\[
 \Prob_K[F_K\ne g|_K]\le C\tau_h.
\]
It has $\delta$-approximate soundness $C$ if every instance admits $g$ such that
\[
 \Prob_K[d(F_K,g|_K)\ge\delta|K|]\le C\tau_h.
\]
Both conclusions use the rejection probability of the exact equality test.
\end{defn}

Write $\Delta(\mathcal F)=\min_g\Prob_K[F_K\ne g|_K]$. The agreement expansion constant is the reciprocal of the best soundness constant. We often use coordinatewise majority as a global assignment: for every $x$, choose a most frequent value of $F_K(x)$ under $\mu_k(\cdot\mid x\in K)$, breaking ties arbitrarily.

We also allow each $F_K$ to belong to a prescribed family $\mathcal C_K\subseteq\Sigma^K$. The \emph{restricted-distance promise} at rank $h$ is that, for every $H\in P_h$, any two distinct assignments in
\[
 \{f|_H:K\supseteq H,\ f\in\mathcal C_K\}
\]
have Hamming distance at least $\theta_h|H|$. This promise concerns the restrictions from all parents of $H$.

\subsection{Layered posets and their walks}
\label{subsec:poset-walks}

For our typical applications, the set $\mathcal K$ corresponds to the top-dimensional faces of a layered poset, which also determines the $K,H$ sampling rule of Section~\ref{subsec:intro-statements}.

Let $P=P_0\cup\cdots\cup P_k$ be a layered poset of subsets of $X$, ordered by containment, with $P_j\subseteq\binom{X}{q_j}$. Suppose every $C_h\in P_h$ contains exactly $d_{j,h}$ faces of $P_j$. A distribution $\mu_k$ on $P_k$ induces the measure
\[
 \mu_j(C_j)=\sum_{C_k\supseteq C_j}\frac{\mu_k(C_k)}{d_{j,k}}.
\]
We assume that these measures are coherent: for $i\le j\le h$, sampling a uniform $j$-face of $C_h$ and then a uniform $i$-face gives a uniform $i$-face of $C_h$. Equivalently,
\[
 \#\{C_j:C_i\subseteq C_j\subseteq C_h\}
 =\frac{d_{j,h}d_{i,j}}{d_{i,h}}.
\]
Down-closed simplicial and Grassmannian complexes have this property.

For a fixed tested layer $h<k$, the earlier setup is recovered by taking $\mathcal K=P_k$, $\mu=\mu_k$, $N=q_k$, and $n=q_h$. The sampled face $H$ has distribution $\mu_h$, and, for $H\subseteq K$,
\[
 D(H\mid K)=\frac1{d_{h,k}},\qquad
 U(K\mid H)=\frac{\mu_k(K)}{d_{h,k}\mu_h(H)},
\]
with both kernels zero otherwise. Thus the down-up walk is $K\to H\to K'$. At stationarity, conditional on $H$, the two parents $K,K'$ are independent samples from $U(\cdot\mid H)$, as in the agreement test.

For any two layers $j<h$, define the up and down operators by
\[
 (U_{j,h}f)(C_j)
 =\frac{\sum_{C_h\supseteq C_j}\mu_h(C_h)f(C_h)}
        {\sum_{C_h\supseteq C_j}\mu_h(C_h)},
 \qquad
 (D_{h,j}u)(C_h)
 =\frac1{d_{j,h}}\sum_{C_j\subseteq C_h}u(C_j).
\]
These are the conditional averages associated with the same sampling rule: $U_{j,h}$ averages over a parent and $D_{h,j}$ over a subface. They are adjoint in the corresponding weighted inner products. The down-up walk $D_{h,j}U_{j,h}$ on $P_h$ is reversible with stationary measure $\mu_h$. Let $\gamma_{j,h}(P)$ be its spectral gap. In particular, $D_{k,h}U_{h,k}$ is the $K\to H\to K'$ walk above, with spectral gap $\gamma_{h,k}(P)$. The following consequence will be useful for comparing conditional averages at different layers.

\begin{claim}\label{claim:contraction}
For $f:P_h\to\mathbb R$,
\[
 \Var_{\mu_j}(U_{j,h}f)
 \le(1-\gamma_{j,h}(P))\Var_{\mu_h}(f).
\]
\end{claim}
\begin{proof}
After subtracting the mean, adjointness gives
\[
 \|U_{j,h}f\|_{\mu_j}^2
 =\langle f,D_{h,j}U_{j,h}f\rangle_{\mu_h}
 \le(1-\gamma_{j,h}(P))\|f\|_{\mu_h}^2.
\]
\end{proof}

The main theorem uses the walks $M_x$ from Section~\ref{subsec:intro-statements}, obtained by conditioning the downward step on $x\in H$. Starting at $K\ni x$, we sample $H\subseteq K$ conditional on containing $x$, and then sample $K'$ from $U(\cdot\mid H)$. If the inclusion probability
\[
 a_x=\Prob[x\in H\mid K]
\]
is independent of the parent $K$ containing $x$, then $M_x$ is reversible with stationary measure $\mu_k(\cdot\mid x\in K)$.

These conditioned walks can be studied in links. For $S\subseteq X$, the link consists of the faces containing $S$, with $S$ removed and with the induced conditional measure. In a simplicial complex this reduces ranks by $|S|$; in a linear Grassmannian we use the quotient by the subspace $S$, reducing dimensions by $\dim S$. In particular, if $P^x$ is the simplicial link of $x$, then $M_x$ is its $(h-1,k-1)$ down-up walk and has spectral gap $\gamma_{h-1,k-1}(P^x)$.

\section{Agreement from polarity}\label{sec:agreement}

\subsection{Exact agreement from regularized polarity}
\label{sec:regularized-core}

We will now prove the main result of the paper, Theorem \ref{thm:intro-regularized}, restated below in its precise form. 
\begin{thm}
\label{thm:regularized-exact}
Let $\mathcal K$ be a family of $N$-element sets with distribution
$\mu$, and let a downward kernel $D(H\mid K)$ sample $n$-element sets
$H\subseteq K$. Suppose $0<n<N$, let $\rho=n/N$, and define $U(K\mid H)$ as the kernel which samples $K$ conditional on $H$. Assume the following.
\begin{enumerate}
\item For each
$K$ and every $x\in K$,
\[
 \Prob_{H\sim D(\cdot\mid K)}[x\in H]=\rho.
\]
\item For every $K$ and $E\subseteq K$, writing $m=|E|$ and
$Z=|E\cap H|$, we have
\begin{equation}\label{eq:regularized-sampler}
 \Ex[Z\mid K]=\rho m,\qquad
 \Var(Z\mid K)\le\rho(1-\rho)m.
\end{equation}
\item For every element $x$ of positive marginal mass, the chain
on top sets containing $x$ that samples $H$ downward conditioned on
$x\in H$ and then samples $K'$ upward from $H$ has spectral gap at
least $\gamma>0$. Its stationary measure is
$\mu(\cdot\mid x\in K)$.
\end{enumerate}
For any finite alphabet $\Sigma$ and local functions $F_K:K\to\Sigma$,
let
\[
 \tau=\Prob_{H,K,K'}[F_K|_H\ne F_{K'}|_H],
\]
where $H$ has its induced marginal and $K,K'$ are independent samples conditional on $H$.

Then the coordinatewise majority function $g$ satisfies
\begin{equation}\label{eq:regularized-bound}
 \Prob_K[F_K\ne g|_K]
 \le\left(\frac8\gamma+
          \frac{64(1-\rho)}{\rho\gamma^2}\right)\tau.
\end{equation}
\end{thm}

For a distribution $\pi$ on assignments, we define its \emph{non-polarity} as
$1-\sum_w\pi(w)^2$, the probability that two independent samples
differ. We are interested in the non-polarity of the distributions $\pi$ of assignments $F_K$ induced on $H$ over $K\supseteq H$, and our main goal is to use non-polarity measure to connect the disagreement probability of local assignments with the majority $g$ and the rejection probability of the agreement test. In the proof, we work with a softer weighted version of non-polarity $\mathcal{E}$ (see (\ref{eq:def-E})), with weights roughly corresponding to the discrepancy of assignments from the global majority. Via spectral expansion, we lower bound $\mathcal{E}$ by a suitable regularized version of the disagreement probability $V$ (between $F_K$ and the majority restricted to $K$). 

Our main motivation for the regularized quantities comes from the following random exposure process. We give each element a rate-one exponential clock and add an independent stopping clock of rate $a>0$. Expose elements in clock order until the first error (discrepancy from the majority) or the stop. If an assignment has $m$ errors, each error is the output with probability $1/(m+a)$, and the exposure finds an error with probability
$m/(m+a)$. To connect to the agreement test using the smaller faces $H$, we run the same exposure only on the elements of $H$, with a modified stopping rate. The random exposure on $H$ defines a local version of the non-polarity $\mathcal{E}$, which we show via the variance bound, to be close to $\mathcal{E}$. Furthermore, the local exposure on $H$ only depends on the restriction of assignments to $H$, and thus for $K,K'$ which agree on $H$, they yield identical first error distribution. This shows an upper bound on $\mathcal{E}$ in terms of the agreement test, which yields the desired result. 

\begin{proof}
Choose $g(x)$ to maximize
$\Prob_{K\mid x\in K}[F_K(x)=g(x)]$, breaking ties arbitrarily.
Put
\[
 e_K(x)=\mathbf1_{F_K(x)\ne g(x)},\qquad
 m_K=\sum_{x\in K}e_K(x).
\]
For $a>0$, define
\[
 f(u)=(u+a)^{-1/2},\qquad
 V=\Ex_K\frac{m_K}{m_K+a}.
\]
Thus $V$ is the probability that the stopped exposure finds an error between $F_K$ and $g$.
For each $x$ and $s\ne g(x)$, define for $K$ containing $x$
\[
 \phi_{x,s}(K)=\mathbf1_{F_K(x)=s}\,f(m_K).
\]
Thus $\phi_{x,s}(K)$ is the square root of the probability that the exposure on $K$ outputs the error witness $(x,s)$.
Since $s$ differs from the chosen majority label, $\Prob_{K\mid x}[F_K(x)=s] \le 1/2$. By an application of Cauchy--Schwarz inequality, 
\[
 \bigl(\Ex_{K\mid x}\phi_{x,s}\bigr)^2
 \le\Prob_{K\mid x}[F_K(x)=s]\,
       \Ex_{K\mid x}\phi_{x,s}^2
 \le\frac12\Ex_{K\mid x}\phi_{x,s}^2.
\]
Consequently
$\Var_{K\mid x}\phi_{x,s}\ge\tfrac12\Ex_{K\mid x}\phi_{x,s}^2$.
The spectral gap for the conditioned walks therefore implies
\begin{equation}\label{eq:regularized-point-energy}
 \Ex_{H,K,K'\mid x\in H}
   (\phi_{x,s}(K)-\phi_{x,s}(K'))^2
 \ge\gamma\Ex_{K\mid x}\phi_{x,s}(K)^2.
\end{equation}

Let
\begin{equation}\label{eq:def-E}
 \mathcal E=\Ex_{H,K,K'}
   \sum_{x\in H}\sum_{s\ne g(x)}
       (\phi_{x,s}(K)-\phi_{x,s}(K'))^2.
\end{equation}
We have
$\Prob[x\in H]=\rho\Prob[x\in K]$. Summing
\eqref{eq:regularized-point-energy} over element-label pairs and using
$\sum_{x\in K,s\ne g(x)}\phi_{x,s}(K)^2=m_K/(m_K+a)$ yields
\begin{equation}\label{eq:regularized-lower}
 \mathcal E\ge\rho\gamma V.
\end{equation}

Let $Z=\sum_{x\in H}e_K(x)$, so that $\Ex[Z\mid K]=\rho m_K$. $Z$ is a random variable that depends on both $K$ and $H$ that is counting the number of errors in $H$ from the assignment on $K$. Run the same exposure on the elements of $H$, with rate-one element clocks and stopping rate $\rho a$. Let $\nu_{K,H}$ be the distribution of the first error witness $(x,F_K(x))$ with $x\in H$, i.e. $(x,F_K(x))$ corresponds to the element with smallest clock time with $F_K(x)\ne g(x)$, with outcome $\bot$ if the stopping clock rings before any error is exposed. For $x\in H$ and $s\ne g(x)$,
\[
 \nu_{K,H}(x,s)=\frac{\mathbf1_{F_K(x)=s}}{Z+\rho a},
 \qquad \nu_{K,H}(\bot)=\frac{\rho a}{Z+\rho a}.
\]
Define
\[
 \psi_{x,s}(K,H)=\sqrt{\rho\,\nu_{K,H}(x,s)}
 =\mathbf1_{F_K(x)=s}\,f(Z/\rho).
\]
Observe that $\psi(K,H)=\phi(K)|_H$ whenever $Z=\rho m_K$. Moreover,
\[
 \|\psi(K,H)\|_2^2 = \sum_{x,s} \psi_{x,s}(K,H)^2 = \rho\sum_{x,s} \nu_{K,H}(x,s)
 =\rho\bigl(1-\nu_{K,H}(\bot)\bigr)
 =\frac Z{Z/\rho+a}\le\rho.
\]
On an accepted pair, $F_K|_H=F_{K'}|_H$, so the local witness distributions $\nu_{K,H}$ and $\nu_{K',H}$, and hence the associated $\psi$, are identical. On a rejected pair, we have $\|\psi(K,H)-\psi(K',H)\|_2^2\le 2\rho$. Hence
\begin{equation}\label{eq:regularized-psi}
 \Ex_{H,K,K'}\|\psi(K,H)-\psi(K',H)\|_2^2\le2\rho\tau.
\end{equation}

Finally, we connect $\phi(K)|_H$ and $\psi(K,H)$. Let
$R(K,H)=\phi(K)|_H-\psi(K,H)$. For $z\ge0$ and $u=z/\rho$,
\begin{align*}
 z(f(m)-f(u))^2
 &=\frac{z(m-u)^2}
   {(m+a)(u+a)(\sqrt{m+a}+\sqrt{u+a})^2}\\
 &\le\frac{\rho(m-u)^2}{(m+a)^2}.
\end{align*}
Using \eqref{eq:regularized-sampler} and
$\Ex[Z/\rho\mid K]=m_K$, we obtain
\begin{align*}
 &\Ex_{H\mid K}\|R(K,H)\|_2^2\\
 &=\Ex_{H\mid K} \sum_{x\in H, s\ne g(x)} \mathbf{1}_{F_K(x)=s} (f(m_K) - f(Z/\rho))^2\\
 &=\Ex_{H\mid K}Z(f(m_K)-f(Z/\rho))^2\\
 &\le\frac{\Var(Z\mid K)}{\rho(m_K+a)^2}
 \le\frac{1-\rho}{a}\frac{m_K}{m_K+a}.
\end{align*}
Therefore
\begin{equation}\label{eq:regularized-residual}
 \Ex_{K,H}\|R(K,H)\|_2^2\le\frac{1-\rho}{a}V.
\end{equation}
Conditional independence of $K,K'$ on $H$ gives
\[
 \Ex[\|R(K,H)-R(K',H)\|_2^2\mid H]
 \le2\Ex[\|R(K,H)\|_2^2\mid H].
\]
Hence, 
\begin{align*}
    \mathcal{E} &= \Ex_{H,K,K'}[\|\phi(K)|_H - \phi(K')|_H\|_2^2] \\
    &= \Ex_{H,K,K'}[\|R(K,H) - R(K',H) + \psi(K,H)-\psi(K',H)\|_2^2] \\
    &\le \frac{4(1-\rho)}{a}V + 4\rho \tau,
\end{align*}
by using \eqref{eq:regularized-psi}, and
\eqref{eq:regularized-residual}. 

Combining with
\eqref{eq:regularized-lower}, we obtain
\[
 \rho\gamma V\le\mathcal E
 \le4\rho\tau+\frac{4(1-\rho)}aV.
\]
Since a nonzero $m_K$ is at least one, $V\ge \Prob_K[m_K>0]/(1+a)$. Then, by choosing $a=8(1-\rho)/(\rho\gamma)$, 
\[
 \Prob_K[m_K>0]\le\frac{8(1+a)}\gamma\tau
 =\left(\frac8\gamma+
         \frac{64(1-\rho)}{\rho\gamma^2}\right)\tau.
\]
\end{proof}

\begin{cor}[Deviation on the test face]
\label{cor:regularized-nonpolarity}
Under the hypotheses of Theorem~\ref{thm:regularized-exact}, we have
\begin{equation}\label{eq:regularized-intersection-deviation}
 \Prob_{K,H}[F_K|_H\ne g|_H]
 \le\left(\frac8\gamma+
          \frac{64(1-\rho)}{\gamma^2}\right)\tau.
\end{equation}
\end{cor}

\begin{proof}
By an application of Jensen's inequality,
\[
 W:=\Ex_{K,H}\frac Z{Z+\rho a}
 \le\Ex_K\frac{\rho m_K}{\rho m_K+\rho a}=V.
\]
Since $Z$ is an integer,
$W\ge\Prob[Z>0]/(1+\rho a)$. Substituting
$V\le8\tau/\gamma$ and $a=8(1-\rho)/(\rho\gamma)$ proves the corollary.
\end{proof}

\subsection{An average Hamming bound}

In this section, we give a simple argument to show that for a typical face $K$, $d(F_K, g|_K)$ is small. We note that for the soundness parameter of the exact agreement test, in contrast to the result in the earlier subsection, this simple argument gives a significantly worse bound (dependent on the size of the top face). We include it as it is useful for approximate agreement results.
\begin{thm}
\label{thm:main-tech}\label{thm:main-tech-d}
Let the top faces have size $q_k$ and the test faces size $q_h$. Suppose that for
each element $x$
\[
 a_x=\Prob_{H\mid K}[x\in H]
\]
is positive and independent of the parent $K$ containing $x$.
Assume that every stationary top--test--top chain conditioned on an element
has spectral gap at least $\gamma>0$, and put $\xi=\min_xa_x$.
For every finite-alphabet assignment, the coordinatewise majority
function $g$ satisfies
\[
 \Ex_Kd(F_K,g|_K)\le\frac{q_h}{\gamma\xi}\tau.
\]
Consequently,
\[
 \Prob_K[F_K\ne g|_K]\le\frac{q_h}{\gamma\xi}\tau,
 \qquad
 \Prob_K[F_K\ne_\delta g|_K]
 \le\frac{q_h}{\gamma\xi\delta q_k}\tau
 \quad(\delta>0).
\]
If $a_x=q_h/q_k$ for every element, then
\[
 \Ex_K\frac{d(F_K,g|_K)}{q_k}\le\frac\tau\gamma.
\]
\end{thm}

\begin{proof}
Let $\nu$ be the distribution of a uniform element of the sampled
$H$, and write
\[
 p_{x,s}=\Prob_{K\mid x}[F_K(x)=s],\qquad
 p_{H,x,s}=\Prob_{K\mid H}[F_K(x)=s].
\]
Apply the spectral gap assumption, we obtain
\[
 \Ex_{H\mid x}\left[1-\sum_s p_{H,x,s}^2\right]
 \ge\gamma\left(1-\sum_s p_{x,s}^2\right)
 \ge\gamma(1-p_{x,g(x)}).
\]
Therefore the expected normalized Hamming disagreement satisfies
\begin{align*}
 D_{\rm Ham}
 &:=\Ex_{H,K,K'}\frac{d(F_K|_H,F_{K'}|_H)}{q_h}\\
 &\ge\gamma\Ex_{x\sim\nu}(1-p_{x,g(x)})
 =\gamma\Ex_{K,H}\frac{d(F_K|_H,g|_H)}{q_h}.
\end{align*}
Since $D_{\rm Ham}\le\tau$ and
\[
 \Ex_{H\mid K}\frac{d(F_K|_H,g|_H)}{q_h}
 =\frac1{q_h}\sum_{x\in K}a_x\mathbf1_{F_K(x)\ne g(x)}
 \ge\frac\xi{q_h}d(F_K,g|_K),
\]
the absolute-error estimate follows. A nonzero Hamming distance is
at least one, and Markov's inequality at threshold $\delta q_k$
gives the desired conclusions.
\end{proof}

\section{Applications}\label{sec:applications}

\subsection{One-sided local spectral expanders}

A pure simplicial complex of rank $k$ is the downward closure of a family $P_k\subseteq\binom{X}{k}$. We equip $P_k$ with a probability distribution and all lower layers with its induced measures. Thus ranks in this paper count vertices. For $0\le i\le k-2$, let $\alpha_i\in[0,1)$ be an upper bound on the second largest eigenvalue of the non-lazy walk on the graph of every $i$-face link.

The within-face sampling assumption of Theorem~\ref{thm:regularized-exact} follows immediately from the hypergeometric variance formula. If $E\subseteq K$, $|E|=m$, and $H$ is a uniform $h$-subset of $K$, then, with $\rho=h/k$,
\[
 \Ex|E\cap H|=\rho m,\qquad
 \Var(|E\cap H|)
 =\rho(1-\rho)m\frac{k-m}{k-1}
 \le\rho(1-\rho)m.
\]
The relation between local spectral expansion and high order walks was developed in~\cite{KM17,KO18,AL20}. Related local-to-global arguments underlie the theory of spectral independence~\cite{ALO24} and its applications to sampling~\cite{chen2021rapid,feng2022rapid}. We use the following local-to-global walk estimate~\cite[Fact A.8 and Theorem A.9]{CLV}.

\begin{thm}[Local-to-global walk bound]\label{thm:local-to-global}
For $1\le h<k$,
\[
 \gamma_{h,k}(P)
 \ge\frac{\sum_{j=h-1}^{k-2}\beta_j}
          {1+\sum_{j=0}^{k-2}\beta_j},
 \qquad
 \beta_j=\prod_{i=0}^{j}\frac{1-\alpha_i}{1+\alpha_i}.
\]
\end{thm}

\begin{cor}\label{cor:one-sided-agreement-expansion}
Suppose $\alpha_i\le a<1$ and $\sum_{i=0}^{k-2}\alpha_i\le A$. Then the exact $(k,h)$-agreement test has soundness
\[
 O_{a,A}\!\left(\frac{k^2}{h(k-h)}\right).
\]
Its $\delta$-approximate soundness is $O_{a,A}(k/(\delta(k-h)))$.
\end{cor}
\begin{proof}
For $0\le t\le a$, $\log((1-t)/(1+t))\ge-c_a t$ for a constant $c_a$. Hence $e^{-c_a A}\le\beta_j\le1$. The same bounds hold for the products in each singleton link. Applying Theorem~\ref{thm:local-to-global} in that link gives
\[
 \gamma_{h-1,k-1}(P^x)
 \ge e^{-c_a A}\frac{k-h}{k-1}.
\]
When $h=1$, the conditioned walk is an independent refresh and has spectral gap $1$. The exact conclusion now follows from Theorem~\ref{thm:regularized-exact} with $\rho=h/k$. The normalized Hamming estimate in Theorem~\ref{thm:main-tech} and Markov's inequality give the approximate conclusion.
\end{proof}

In particular, if $\alpha_i\le c/k$ for a fixed $0<c<1$, the exact soundness is $O_{c,\varepsilon}(1)$ for $\varepsilon k\le h\le(1-\varepsilon)k$, and $O_c(k)$ for adjacent layers $h=k-1$. For Ramanujan complexes with local spectral bound $O(q^{-1/2})$, the same conclusions hold when $q\ge Ck^2$ for a sufficiently large absolute constant $C$~\cite{DK17}.

\subsection{Full bases of matroids}

The independence complex of a matroid has one-sided local spectral bounds $\alpha_i=0$~\cite{ALOV19}. As a consequence of Theorem \ref{thm:regularized-exact}, we obtain the following corollary. 

\begin{cor}[Agreement on matroid bases]
\label{cor:regularized-matroid}\label{cor:agreement-expansion-matroids}
Let $\mu_k$ be the uniform distribution on the bases of a rank-$k$ matroid, where $k\ge2$. For $1\le h<k$, the exact $(k,h)$-agreement test has soundness at most
\begin{equation}\label{eq:regularized-matroid}
 \frac{8(k-1)}{k-h}+\frac{64(k-1)^2}{h(k-h)}.
\end{equation}
In particular, its soundness is $O_\varepsilon(1)$ for $\varepsilon k\le h\le(1-\varepsilon)k$, and at most $72(k-1)$ for $h=k-1$.
\end{cor}
\begin{proof}
For every nonloop $x$, the link is the independence complex of the contraction by $x$. Its $(h-1,k-1)$ down-up walk has spectral gap at least $(k-h)/(k-1)$~\cite{ALOV19,AL20}. Substituting this bound and $\rho=h/k$ in Theorem~\ref{thm:regularized-exact} gives~\eqref{eq:regularized-matroid}.
\end{proof}

\subsection{Grassmannian posets}\label{subsec:grassmannians}

We now apply Theorem~\ref{thm:regularized-exact} to Grassmannian posets. Let $P$ be a down-closed family of linear or affine subspaces over $\mathbb F_q$, with all maximal spaces of dimension $k$. We put a distribution on the $k$-spaces and sample a uniform $h$-space inside each of them, where $1\le h<k$. The alphabet $\Sigma$ is arbitrary.

First consider assignments on projective points in the linear case and on affine points in the affine case. Recall that
\[
 n_j=\frac{q^j-1}{q-1}
\]
is the number of projective points in a linear $j$-space. Thus the sizes of the top and tested sets are $(N,n)=(n_k,n_h)$ in the projective case and $(N,n)=(q^k,q^h)$ in the affine case. As verified in Section~\ref{subsec:intro-statements}, uniform subspace sampling satisfies both sampling assumptions of the main theorem.

\begin{cor}[Exact agreement on Grassmannians]\label{cor:grassmannian-exact}
Suppose the down-up walk conditioned on each projective point, or each affine point in the affine case, has spectral gap at least $\gamma>0$. Then the exact $(k,h)$-agreement test has soundness at most
\begin{equation}\label{eq:grassmannian-exact}
 C(\rho,\gamma)=\frac8\gamma+\frac{64(1-\rho)}{\rho\gamma^2},
 \qquad
 \rho=\begin{cases}
 \dfrac{q^h-1}{q^k-1},&\text{in the projective case},\\[4pt]
 q^{h-k},&\text{in the affine case}.
 \end{cases}
\end{equation}
For assignments on all vectors of linear spaces, assume also that the unconditioned down-up walk has spectral gap $\gamma_0>0$. The soundness is then at most $C(\rho,\gamma)+1/\gamma_0$, with the projective value of $\rho$.
\end{cor}
\begin{proof}
The projective and affine conclusions follow directly from Theorem~\ref{thm:regularized-exact}. For assignments on all vectors, group the labels on the $q-1$ nonzero vectors of each projective point into one label in $\Sigma^{q-1}$. The rejection probability of this grouped test is at most $\tau$. Applying the projective result gives a global assignment on the nonzero vectors. Choose its value at the origin by majority. If $(p_s)_{s\in\Sigma}$ is the distribution of $F_K(0)$, the Poincar\'e inequality gives
\[
 \Prob_K[F_K(0)\ne g(0)]
 =1-\max_s p_s
 \le1-\sum_s p_s^2
 \le\frac\tau{\gamma_0}.
\]
Adding this probability to the projective bound proves the all-vector conclusion.
\end{proof}

For fixed positive spectral gaps, the soundness is $O(q^{k-h})$. We next obtain the required spectral gaps from one-sided local spectral expansion. For every $i$-dimensional space $S$, consider the graph on one-dimensional spaces in its quotient link, with the edge between distinct $L,L'$ weighted proportionally to the induced measure of $L+L'$. Let $\alpha_i\ge0$ bound the second largest eigenvalue of all these graphs. We use the following estimate from the companion paper~\cite{compa}.

\begin{thm}[Grassmannian walk bound, \cite{compa}]\label{thm:grassmannian-walk}
Let $P$ be a linear Grassmannian poset and $\alpha=\max_{0\le i\le k-2}\alpha_i$. If $\alpha<c/k$, where $c>0$ is a sufficiently small absolute constant, then for $1\le h<k$,
\[
 \gamma_{h,k}(P)
 \ge1-\left(\frac1{q-1/q}+k\alpha\right)^{k-h}.
\]
\end{thm}

Applying this estimate in $P$ gives a constant spectral gap for the unconditioned walk. For $h\ge2$, applying it in each one-dimensional quotient gives a constant spectral gap for every walk conditioned on a projective point. When $h=1$, the conditioned walk is an independent refresh and has spectral gap $1$. We therefore obtain the following consequences.

\begin{cor}\label{cor:Grassmannian-(k,k-1)}\label{cor:Grassmannian-(k,k/2)}
Under the assumptions of Theorem~\ref{thm:grassmannian-walk}, the exact $(k,k-1)$-agreement test has soundness $O(q)$. For even $k$, the exact $(k,k/2)$-agreement test has soundness $O(q^{k/2})$. These conclusions hold for assignments on projective points or on all vectors.
\end{cor}
\begin{proof}
Apply Corollary~\ref{cor:grassmannian-exact} with the constant spectral gaps above.
\end{proof}

We also record a simple estimate for restrictions to linear subspaces, used in Section~\ref{sec:local-distance}.
\begin{lem}[Linear subspace avoidance]\label{lem:xi-grass}
Let $0\le\delta\le1$ and $0\le h\le k$. If $A$ is a subset of a linear $k$-space with $|A|\le(1-\delta)q^k$, then a uniform $h$-subspace $H$ satisfies
\[
 \Prob[H\subseteq A]\le(1-\delta)^h.
\]
\end{lem}
\begin{proof}
If $0\notin A$, the probability is zero. Otherwise expose a uniform flag $V_0<\cdots<V_h$. Conditional on $V_i\subseteq A$, choose a uniform $x\notin V_i$ and put $V_{i+1}=\operatorname{span}(V_i,x)$. The event $V_{i+1}\subseteq A$ requires $x\in A$, which has probability
\[
 \frac{|A|-q^i}{q^k-q^i}\le1-\delta.
\]
Iterating proves the claim.
\end{proof}

\subsection{Approximate agreement}\label{sec:approximate-agreement}

We can obtain substantially better soundness if errors on a small fraction of each top face are allowed. The next corollary gives constant soundness when this fraction is at least the inverse size of a tested face, provided the spectral gaps are bounded below by a positive constant.

\begin{cor}\label{cor:approximate-two-design}
Under the hypotheses of Theorem~\ref{thm:regularized-exact}, let $g$ be its majority function. Recall that the top faces have $N$ elements, the tested faces have $n$ elements, and $\rho=n/N$. Put
\[
 A_h=\frac8\gamma+\frac{64(1-\rho)}{\gamma^2}.
\]
For every $0<\delta\le1$,
\begin{equation}\label{eq:approximate-two-design}
 \Prob_K[d(F_K,g|_K)\ge\delta N]
 \le\left(1+\frac{1-\rho}{\delta n}\right)A_h\tau.
\end{equation}
\end{cor}
\begin{proof}
Fix $K$ and write $m=d(F_K,g|_K)$, $Z=d(F_K|_H,g|_H)$, and $\lambda=\rho m$. By the sampling assumptions,
\[
 \Ex[Z\mid K]=\lambda,\qquad
 \Ex[Z^2\mid K]\le\lambda^2+(1-\rho)\lambda.
\]
Thus, whenever $m\ge\delta N$, Cauchy--Schwarz gives
\[
 \Prob[Z>0\mid K]
 \ge\frac{\lambda}{\lambda+1-\rho}
 \ge\frac{\delta n}{\delta n+1-\rho}.
\]
Averaging over $K$ and applying Corollary~\ref{cor:regularized-nonpolarity}, we obtain
\[
 \frac{\delta n}{\delta n+1-\rho}\Prob_K[m\ge\delta N]
 \le\Prob_{K,H}[Z>0]\le A_h\tau,
\]
which proves the result.
\end{proof}

For fixed positive spectral gaps, the $\delta$-approximate soundness is therefore $O_c(1)$ whenever $\delta\ge c/n$, for any fixed $c>0$. In particular, it suffices that
\[
 \delta\ge\frac c h\quad\text{for simplicial complexes},\,\,
 \delta\ge cq^{-h}\quad\text{for affine Grassmannians},\,\,
 \delta\ge\frac c{n_h}\quad\text{for projective Grassmannians}.
\]

For assignments on all vectors of a linear space, apply the corollary to the assignments on projective points constructed in the proof of Corollary~\ref{cor:grassmannian-exact}. Define $g$ by the resulting majority labels on the nonzero vectors, and choose $g(0)$ arbitrarily. If $d(F_K,g|_K)\ge\delta q^k$ and $\delta q^k\ge2$, then the number of projective points whose labels disagree is at least
\[
 \frac{\delta q^k-1}{q-1}\ge\frac\delta2 n_k.
\]
Applying the corollary with tolerance $\delta/2$ gives
\begin{equation}\label{eq:approximate-vector}
 \Prob_K[d(F_K,g|_K)\ge\delta q^k]
 \le\left(1+\frac{2(1-\rho)}{\delta n_h}\right)A_h\tau,
 \qquad \rho=\frac{q^h-1}{q^k-1}.
\end{equation}
Thus the same threshold $\delta\ge c/n_h$, with fixed $c\ge1$, gives constant approximate soundness for all-vector assignments. Under the assumptions of Theorem~\ref{thm:grassmannian-walk}, the sufficient tolerances are therefore of order $q^{2-k}$ for $h=k-1$ and $q^{1-k/2}$ for $h=k/2$.

\section{Agreement under a local distance guarantee}
\label{sec:local-distance}

In this section, we prove a general agreement theorem under a local distance guarantee. The main idea is that, under this setup, we can pass from the tested layer to an auxiliary layer on which a nonempty disagreement is easier to detect. For linear Grassmannians, this improves the soundness from $O(q^{k-h})$ to $O(q)$, independently of the separation between the two layers.

Fix the top layer $k$ of the poset. For each rank $s<k$, define
\begin{equation}\label{eq:xi}
 \xi_{k,s}
 =\min_{\substack{C_k\in P_k\\ f_1\ne f_2\in\{0,1\}^{C_k}}}
   \Prob_{C_s\subseteq C_k}[f_1|_{C_s}\ne f_2|_{C_s}].
\end{equation}
Equivalently,
\begin{equation}\label{eq:xi-elements}
 \xi_{k,s}
 =\min_{\substack{C_k\in P_k\\ \varnothing\ne E\subseteq C_k}}
    \Prob_{C_s\subseteq C_k}[C_s\cap E\ne\varnothing]
 =\min_{\substack{C_k\in P_k\\ x\in C_k}}
    \Prob_{C_s\subseteq C_k}[x\in C_s].
\end{equation}
For uniform subface sampling in downward-closed posets,
\[
 \xi_{k,s}=\begin{cases}
 s/k,&\text{for simplicial complexes},\\[2pt]
 (q^s-1)/(q^k-1),&\text{for linear Grassmannians},\\[2pt]
 q^{s-k},&\text{for affine Grassmannians}.
 \end{cases}
\]

Now fix the tested layer $h$. For $0<\kappa\le1$, let $R(\kappa)$ consist of $r<k$ with the following property: for every $C_r\subseteq C_k$ and every $E\subseteq C_k$ satisfying $|E\cap C_r|\ge\kappa|C_r|$,
\[
 \Prob_{C_h\subseteq C_k}[C_h\cap E\ne\varnothing]\ge\frac12.
\]

\begin{thm}\label{thm:agreement-code}
Let $r\in R(\kappa)$ with $\xi_{k,r}>0$. Suppose that, for each $s\in\{h,r\}$, the local assignments admit a global assignment $g_s$ satisfying
\begin{equation}\label{eq:transfer-deviation}
 \Prob_{C_k,C_s\subseteq C_k}[F_{C_k}|_{C_s}\ne g_s|_{C_s}]
 \le A_s\tau_s,
\end{equation}
where $\tau_s$ is the rejection probability of the $(k,s)$ test. If distinct restrictions of the local assignments to every $r$-face have relative distance at least $\theta_r\ge2\kappa$, then
\begin{equation}\label{eq:xi-transfer}
 \Delta\le\frac{4A_hA_r}{\xi_{k,r}}\tau_h.
\end{equation}
\end{thm}
\begin{proof}
Let
\[
 Q=\left\{C_k:\Prob_{C_h\subseteq C_k}
 [F_{C_k}|_{C_h}\ne g_h|_{C_h}]\ge\frac12\right\}.
\]
Then $\mu_k(Q)\le2A_h\tau_h$. If $C_k\notin Q$, the definition of $R(\kappa)$ gives
\[
 d(F_{C_k}|_{C_r},g_h|_{C_r})<\kappa|C_r|
 \qquad\text{for every }C_r\subseteq C_k.
\]
Thus any $C_k,C_k'$ outside $Q$ have restrictions at distance less than $2\kappa|C_r|$ on their common $r$-face. The distance guarantee forces these restrictions to agree. As such,
\[
 \tau_r\le2\mu_k(Q)\le4A_h\tau_h.
\]
On the other hand, whenever $F_{C_k}\ne g_r|_{C_k}$, a random $r$-face detects this disagreement with probability at least $\xi_{k,r}$. Therefore
\begin{align*}
 \xi_{k,r}\Delta
 &\le\xi_{k,r}\Prob_{C_k}[F_{C_k}\ne g_r|_{C_k}]\\
 &\le\Prob_{C_k,C_r\subseteq C_k}[F_{C_k}|_{C_r}\ne g_r|_{C_r}]
 \le A_r\tau_r.
\end{align*}
Combining the two bounds proves the theorem. 
\end{proof}

The deviation bounds~\eqref{eq:transfer-deviation} follow from our non-polarity argument. Suppose that, at each $s\in\{h,r\}$, the sampling assumptions of Theorem~\ref{thm:regularized-exact} hold with $\rho_s=|C_s|/|C_k|$, and all element-conditioned down-up walks have spectral gap at least $\gamma_s>0$. Then $\xi_{k,s}=\rho_s$, and Corollary~\ref{cor:regularized-nonpolarity} gives
\begin{equation}\label{eq:transfer-spectral}
 A_s=\frac8{\gamma_s}+\frac{64(1-\rho_s)}{\gamma_s^2}.
\end{equation}
Thus Theorem~\ref{thm:agreement-code} gives an explicit soundness bound in terms of $\xi_{k,r}$ and the two spectral gaps. When the spectral gaps are bounded below by a positive constant, it gives soundness $O(1/\xi_{k,r})$, compared with $O(1/\xi_{k,h})$ from applying the deviation bound directly at the tested layer. We may minimize~\eqref{eq:xi-transfer} over all auxiliary ranks in $R(\kappa)$ satisfying the distance guarantee.

We now apply the theorem to linear Grassmannians. The next two concentration bounds verify the hitting property defining $R(\kappa)$; we then connect the distance guarantee from the tested layer to the auxiliary layer. In the following lemmas, $C_j$ denotes a linear space of dimension $j$, and assignments are defined on all its vectors.

\begin{lem}\label{lem:concentration}
Let $0<\kappa\le1$, let $C_r\subseteq C_k$, and put $t=h+r-k\ge0$. Suppose $E\subseteq C_r$ has size at least $\kappa q^r$. Then
\[
 \Prob_{C_h\subseteq C_k}[C_h\cap E=\varnothing]\le(1-\kappa)^t.
\]
In particular, $r\in R(\kappa)$ if $h+r-k\ge\lceil\log(2)/\kappa\rceil$.
\end{lem}
\begin{proof}
The intersection $C_h\cap C_r$ has dimension at least $t$. By invariance under linear automorphisms of $C_r$, a uniform $t$-subspace of this intersection is uniform among the $t$-subspaces of $C_r$. Avoidance by $C_h$ implies avoidance by that subspace. The flag argument of Lemma~\ref{lem:xi-grass}, applied inside $C_r$, bounds this probability by $(1-\kappa)^t\le e^{-\kappa t}$.
\end{proof}

\begin{lem}\label{lem:hitting-moment}
Let $0<\kappa\le1$ and $E\subseteq C_r\subseteq C_k$, with $|E|\ge\kappa q^r$, and put $t=r+h-k\ge0$. Then
\[
 \Prob_{C_h\subseteq C_k}[C_h\cap E=\varnothing]
 \le\frac{q-1}{q-1+\kappa(q^t-1)}.
\]
In particular, $r\in R(\kappa)$ if $\kappa(q^t-1)\ge q-1$.
\end{lem}
\begin{proof}
If $0\in E$, every linear subspace meets $E$. Otherwise, for $Y=|C_h\cap E|$,
\[
 \Ex Y=|E|\frac{q^h-1}{q^k-1}\ge\kappa(q^t-1).
\]
Covariance terms corresponding to vectors on different projective lines are nonpositive. Each vector has at most $q-2$ other vectors on its line, so
\[
 \Ex Y^2\le(\Ex Y)^2+(q-1)\Ex Y.
\]
Cauchy--Schwarz gives $\Prob[Y>0]\ge(\Ex Y)^2/\Ex Y^2$, proving the claim.
\end{proof}

\begin{lem}\label{lem:distance-promotion}
Suppose distinct restrictions of the local assignments to every $h$-face have relative distance at least $\theta$. Then their distinct restrictions to every $r$-face, $r\ge h$, have relative distance at least $\theta-q^{1-h}$.
\end{lem}
\begin{proof}
Let $D\subseteq C_r$ be a nonempty disagreement set. If $D$ contains a nonzero vector $x$, sample an $h$-subspace $H\subseteq C_r$ containing $x$. At most $q$ vectors of $D$ lie in $\operatorname{span}(x)$, and each other vector has conditional inclusion probability
\[
 \frac{q^h-q}{q^r-q}\le q^{h-r}\qquad(r>1).
\]
Every such restriction differs, so
\[
 \theta\le\Ex\frac{|D\cap H|}{q^h}
 \le\frac{|D|}{q^r}+q^{1-h}.
\]
The case $r=h=1$ follows directly from the hypothesis. If $D=\{0\}$, unconditioned sampling gives the same bound with additive term $q^{-h}$.
\end{proof}

\begin{cor}\label{cor-agreement-code}
Let $P$ be a down-closed linear Grassmannian with assignments on all vectors. Suppose distinct restrictions to each tested $h$-face have relative distance at least $\theta$, where $0<\theta\le1$, and
\[
 q^{h-1}\ge1+\frac{8(q-1)}\theta,
 \qquad 1\le h<k.
\]
For each $s\in\{h,k-1\}$, assume that the top--$s$--top walk and every such walk conditioned on a projective point have spectral gaps at least $\gamma_*>0$. Then the exact $(k,h)$ test has soundness $O_{\gamma_*}(q)$.
\end{cor}
\begin{proof}
For each $s\in\{h,k-1\}$, group the nonzero vectors on each projective line as in Corollary~\ref{cor:grassmannian-exact}. Apply Corollary~\ref{cor:regularized-nonpolarity} to these grouped assignments, and choose the value at the origin by majority. The origin estimate in Corollary~\ref{cor:grassmannian-exact} adds at most $\tau_s/\gamma_*$. Hence~\eqref{eq:transfer-deviation} holds with
\[
 A_s\le\frac9{\gamma_*}+\frac{64}{\gamma_*^2}=:A.
\]

Set $\kappa=\theta/8$ and $r=k-1$. The assumed threshold gives $q^{1-h}\le\theta/2$, so Lemma~\ref{lem:distance-promotion} gives $\theta_r\ge\theta/2>2\kappa$. It also gives $\kappa(q^{h-1}-1)\ge q-1$, so Lemma~\ref{lem:hitting-moment} implies $r\in R(\kappa)$. Finally,
\[
 \xi_{k,k-1}^{-1}=\frac{q^k-1}{q^{k-1}-1}\le q+1.
\]
Theorem~\ref{thm:agreement-code} therefore gives
\[
 \Delta\le4A^2(q+1)\tau_h=O_{\gamma_*}(q)\tau_h.
\]
\end{proof}

The argument in Lemma~\ref{lem:concentration} also proves the conclusion under the simpler sufficient condition $h-1\ge\lceil8\log(2)/\theta\rceil$.

\end{document}